\RequirePackage{fix-cm}
\documentclass[twocolumn,epjc3]{svjour3}  
\smartqed  % flush right qed marks, e.g. at end of proof
\RequirePackage{graphicx}
\usepackage{rotating}
\usepackage{lipsum}
\journalname{Eur. Phys. J. C}
\begin{document}

\title{Forecasting constraints on the baryon mass fraction in the IGM from fast radio bursts and type Ia supernovae}

%\titlerunning{Forecasting FRB constraints on $f_{IGM}$}        % if too long for running head

\author{Thais Lemos\thanksref{e1,addr1}
        \and
        Rodrigo Gon\c{c}alves\thanksref{e2,addr2} %etc
        \and
        Joel Carvalho\thanksref{e3,addr1}
        \and
        Jailson Alcaniz\thanksref{e4,addr1}
}

\thankstext{e1}{e-mail: thaislemos@on.br (corresponding author)}
\thankstext{e2}{e-mail: rsg\_goncalves@ufrrj.br}
\thankstext{e3}{e-mail: jcarvalho@on.br}
\thankstext{e4}{e-mail: alcaniz@on.br}


\institute{Observat\'orio Nacional, Rio de Janeiro - RJ, 20921-400, Brazil \label{addr1}
           \and
           Departamento de F\'{\i}sica, Universidade Federal Rural do Rio de Janeiro, Serop\'edica - RJ, 23897-000, Brazil \label{addr2}
           %\and
           %\emph{Present Address:} if needed\label{addr3}
}

\date{Received: date / Accepted: date}

\maketitle

\begin{abstract}

 Fast Radio Bursts (FRBs) are millisecond transient radio events with a high energy. By identifying the origin of the burst, it is possible to measure the redshift of the host galaxy, which can be used to constrain cosmological and astrophysical parameters and test aspects of fundamental physics when combined with the dispersion measure ($DM$). However, some factors limit the cosmological application of FRBs: (i) the poor modelling of the fluctuations in the $DM$ due to spatial variation in the cosmic electrons density; (ii) the fact that the fraction of baryon mass in the intergalactic medium ($f_{IGM}$) is degenerated with some cosmological parameters; (iii) the limited knowledge about host galaxy contribution ($DM_{host}$). In this work, we investigate the impact of different redshift distribution models of FRBs to constrain the baryon fraction in the IGM and host galaxy contribution. We use a cosmological model-independent method developed in previous work \cite{Lemos2023} to perform the analysis and combine simulated FRB data from Monte Carlo simulation and supernovae data. We assume four distribution models for the FRBs: gamma-ray bursts (GRB), star formation rate (SFR), uniform and equidistant (ED). Also, we consider samples with $N = 15, 30, 100$ and $500$ points and different values of the fluctuations of electron density in the $DM$, $\delta = 0, 100, 200, 400, 230\sqrt{z}$ pc/cm$^{3}$. Our analysis shows that all the distribution models present consistent results within $2\sigma$ for the free parameters $f_{IGM}$ and $DM_{host,0}$ and highlights the crucial role of $DM$ fluctuations in obtaining more precise  measurements.

\keywords{Cosmology \and Fast Radio Burst \and Monte Carlo Simulation \and Intergalactic Medium \and Baryon Fraction}
\end{abstract}

\section{Introduction}
\indent

Fast Radio Bursts (FRBs) are high-energy transient events with a millisecond duration and radio frequency range of a few hundred to a few thousand MHz  \cite{Petroff2022,Thornton2013,Petroff2015,Petroff2016,Platts2019}. In the past years, some models have been proposed to explain the origin of the burst, but the physical mechanism responsible for it is still in debate~\cite{Petroff2021}. However, the large observed dispersion measure ($DM$) above that of the Milk Way suggests an extragalactic or cosmological origin for the FRBs \cite{Dolag2015}. Since  their first discovery by Parkes Telescope in 2007 \cite{Lorimer2007}, more than one hundred FRBs have been detected thanks to new telescopes, such as e.g. the Canadian Hydrogen Intensity Mapping Experiment (CHIME, \cite{CHIME}).

It is a common understanding that some of their observational properties must be better understood to explore the full potential of these objects in both astrophysical and cosmological contexts. For instance, due to the spatial variation in cosmic electron distribution, the density fluctuations in the dispersion measure ($DM$) need to be better determined~\cite{Takahashi2021}. Another limitation is the poor knowledge about the host galaxy contribution of the FRBs ($DM_{host}$), which depends on many factors such as the galaxy type, the relative orientation between the FRB source with respect to the host as well as the mass of the host galaxy ~\cite{Xu2015}. The redshift evolution of $DM_{host}$ remains unknown and previous works studied different functions as such as simple log-normal form with median value of $100$ pc $\rm{cm^{-3}}$ \cite{Walker2020}, as well as a normal or log-normal distribution with a median value as free parameter in the range $20-200$ pc/$\rm{cm^{3}}$ \cite{Macquart2020}, among others.

When the origin of the burst is confirmed, the galaxy host can be identified, and the redshift of the event can be measured directly. In this situation, the dispersion measure can be combined with the redshift to obtain the $DM-z$ relation \cite{Deng2014}. From these relations, one can use FRBs to probe the anisotropic distribution of baryon matter in Universe \cite{Lin2021}, to test the weak equivalence principle \cite{Reischke2023} or to constrain cosmological parameters \cite{Walters2018,Wei2018}, such as the Hubble constant \cite{Wu2020,Hagstotz,Wu2021} and the baryon mass fraction in the intergalactic medium ($f_{IGM}$) \cite{Li2019,Wei2019,Li2020}.

An interesting aspect regarding the $f_{IGM}$  is the possibility of its variation with respect to redshift. In \cite{Shull2012}, the authors found  $f_{IGM} \approx 0.82$ at $z \geq 0.4 $, while in \cite{Meiksin2009} the authors estimated $f_{IGM} \approx 0.9$ at $z \geq 1.5$. More recently, in a previous communication~\cite{Lemos2023}, we used a cosmological model-independent method to constrain $f_{IGM}$, assuming both constant and time-dependent parameterizations, and found that the time-evolution of $f_{IGM}$ depends strongly on the $DM$ fluctuations due to the spatial variation in cosmic electron density. Among all the previously parameters mentioned, here we focus mainly on $DM_{host}$ and $f_{IGM}$.

One issue when studying FRBs in cosmology is the identification of the host galaxy, and although many events have been observed in the sky, only a few FRBs in the literature are
well localized, with the correspondent redshift \cite{Petroff2019}. The current FRBs sample is not large enough to perform robust statistical analysis, %which makes numerical simulations necessary. 
but instruments are being built to localize FRBs in the next few years. Among these are the coherent upgrade CRACO system \cite{James2022} of Australian Square Kilometre Array Pathfinder (ASKAP), the Canadian Hydrogen Intensity Mapping Experiment (CHIME) outriggers \cite{CHIME} and SKA1-Mid \cite{SKA}. While ASKAP/CRACO is expected to localize $\sim 100$ FRBs per year, the number for CHIME/FRB is $\sim 500$ FRBs per year. 

In this context, understanding the constraining power of the upcoming observations through numerical simulations is, therefore, an important and necessary task. However, to perform
such simulations, it is crucial to determine the redshift distribution of the FRBs. As the origin of them is unknown, it is necessary to combine astrophysical assumptions with numerical simulations to obtain such functions. The literature has explored distributions based on general aspects, such as star formation history/rate~\cite{Bhattacharya2021} or by assuming a specific astrophysical origin, such as gamma-ray bursts~\cite{Zhang2020}. For a general analysis of the possible distributions, we refer the reader to \cite{Qiang2021} and references therein.

%%------------------------------------------------------  TABLE  ------------------------------------------------------

\begin{table*}
\centering
\caption{A list of FRB with known host galaxies.}
\begin{tabular}{ c  c  c  c  c c}
\hline
Name & Redshift $z$ & $DM_{MW, ISM}$ & $DM_{obs}$ & $\sigma_{obs}$ & Reference\\
 &  & [pc/cm$^{3}$] & [pc/cm$^{3}$] & [pc/cm$^{3}$] & \\
\hline
FRB 180916B & 0.0337  & 200.0 & 348.8  & 0.2 &\cite{FRB180916}\\
FRB 201124A & 0.098   & 123.2 & 413.52 & 0.5 &\cite{FRB201124}\\
FRB 190608B & 0.1178  & 37.2  & 338.7  & 0.5 &\cite{FRB190608}\\
FRB 200430A & 0.16    & 27.0    & 380.25 & 0.4 &\cite{FRB190523_2}\\
FRB 121102A & 0.19273 & 188.0 & 557.0  & 2.0 &\cite{FRB121102}\\
FRB 191001A & 0.234   & 44.7  & 506.92 & 0.04 &\cite{FRB190523_2}\\
FRB 190714A & 0.2365  & 38.0    & 504.13 & 2.0 &\cite{FRB190523_2}\\
FRB 20191228A & 0.2432 & 33.0 & 297.5 & 0.05 & \cite{FRB20191228}\\
FRB 190102C & 0.291   & 57.3  & 363.6  & 0.3 &\cite{FRB190102}\\
FRB 180924B & 0.3214  & 40.5  & 361.42 & 0.06 &\cite{FRB180924}\\
%FRB 20180301A & 0.3305 &152.0 & 536.0 & 8.0 & \cite{FRB20191228}\\
FRB 20200906A & 0.3688 & 36.0 & 577.8 & 0.02 & \cite{FRB20191228}\\
FRB 190611B & 0.378   & 57.83 & 321.4  & 0.2 &\cite{FRB190523_2}\\
FRB 181112A & 0.4755  & 102.0 & 589.27 & 0.03 &\cite{FRB181112}\\
FRB 190711A & 0.522   & 56.4  & 593.1  & 0.4 &\cite{FRB190523_2}\\
%FRB 190614D & 0.6     & 83.5  & 959.2  & 5.0 &\cite{FRB190614}\\
FRB 190523A & 0.66    & 37.0  & 760.8  & 0.6 &\cite{FRB190523_1,FRB190523_2}\\
\hline
%\multicolumn{6}{p{2.5cm}}{\,} 
\end{tabular}
\label{tab:FRB}
\end{table*}

In this work, we investigate the impact of different FRB redshift distributions and the number of FRB events on the constraints of $DM_{host}$ and $f_{IGM}$ through Monte Carlo simulations. The redshift distributions are defined from different astrophysical and cosmological assumptions, and we also consider the role of $DM$ fluctuations on the $DM_{host}$ and $f_{IGM}$ estimates. We obtain the mass of baryon fraction in the IGM model-independently as presented in \cite{Lemos2023}, where  FRBs data from Monte Carlo simulated data are combined with type Ia supernovae (SNe) observations. Our results clearly show the crucial role of the $DM$ fluctuations in more precisely determining the cosmological parameters from FRBs observations.

We organized this paper as follows: Sec. \ref{sec:frb} briefly discusses FRBs properties and the main quantities. The data set used and the methodology applied are described in Sec. \ref{sec:data}. Our simulations and results are presented in Secs. \ref{sec:simulations} and \ref{sec:results}, respectively. We end the paper in Sec.  \ref{sec:conclusions} by presenting our main conclusions.

%%-----------------------------------------------------------------------------
\section{FRB Properties}\label{sec:frb}
\indent

The FRB's photons interact with the free electrons in the medium from the host galaxy to the observer on Earth. These interactions result in a change in the frequency of the
pulse, thereby causing a delay in its arrival time. The time delay is proportional to $DM$ and can be written in terms of others components \cite{Macquart2020,Gao2014}
\begin{equation}
\label{DMobs}
    DM_{obs} (z) = \sum_{i}{DM_{i}(z)} \; %\rm{MW,ISM}} + DM_{\rm{MW,halo}} + DM_{host}(z) + DM_{\rm{IGM}}(z) \;,
\end{equation}
where $i = {{MW,ISM}}$; ${host}$; ${{IGM}}$; ${{MW,halo}}$ and are the contributions from the Milky Way interstellar medium (ISM), the host galaxy, the intergalactic medium and the Milky Way halo, respectively.

The term $DM_{\rm{MW, ISM}}$ can be obtained using Galactic electron density models from pulsar observations \cite{Taylor1993,Cordes2002,Yao2017} whereas the halo contribution is not well constrained yet, and therefore, we follow \cite{Macquart2020} and assume $DM_{MW,halo} = 50$ pc/cm$^{3}$. The host galaxy contribution can be written as
\begin{equation}
\label{host}
DM_{host}(z) = \frac{DM_{host,0}}{1+z},
\end{equation}
where the $(1+z)$ factor accounts for the cosmic dilation \cite{Deng2014,Ioka2003}. The host galaxy contribution in the source frame ($DM_{host,0}$) is a poorly known parameter and depends on some factors, such as the type of galaxy and the inclination angle of the host galaxy.  Therefore, in our analysis $DM_{host,0}$ will be treated as a free parameter.

The IGM contribution depends on the redshift and can be written as \cite{Deng2014}
\begin{equation}
\label{IGM}
DM_{IGM}(z)=\frac{3c\Omega_{b}H_{0}^{2}}{8\pi Gm_{p}} \int_{0}^{z} \frac{(1+z')f_{IGM}(z')\chi(z')}{H(z')}dz',
\end{equation}
where $c$,  $\Omega_{b}$, $H_{0}$, $G$, $m_{p}$, $f_{IGM}(z)$, $H(z)$ are, respectively, the speed of light, the present-day baryon density parameter, the Hubble constant, the gravitational constant, the proton mass, the baryon fraction in the IGM and the Hubble parameter at redshift $z$. Also, $\chi(z) = Y_{H}\chi_{e,H}(z) + Y_{He}\chi_{e,He}(z)$ is the free electron number fraction per baryon, in which $Y_{H} = 3/4$ and $Y_{He} = 1/4$ are the mass fractions of hydrogen and helium, respectively, while $\chi_{e,H}(z)$ and $\chi_{e,He}(z)$ are the ionization fractions of hydrogen and helium, respectively. The hydrogen and helium are fully ionized at $z < 3$ \cite{Meiksin2009,Becker2011}, so that we have $\chi_{e,H}(z) = \chi_{e,He}(z) = 1$. 

In \cite{Lemos2023}, we presented a cosmological model-independent method, which solves the $DM_{IGM}$ integral above by parts, identifying one of the terms as the luminosity distance ($d_{L}$). We also considered two parameterizations of the baryon fraction  in terms of the redshift: a constant case, $f_{IGM} (z) = f_{IGM,0}$  and a time-dependent case, $f_{IGM} (z) = f_{IGM,0} + \alpha z/(1+z)$. For simplicity, in the present paper we consider only the constant case, for which Eq. (\ref{IGM}) can be written as 
\begin{equation}\label{constant}
    DM_{IGM}(z) = A f_{IGM,0} \left[\frac{d_{L}(z)}{c} - \int_{0}^{z} \frac{d_{L}(z')}{(1+z')c} dz' \right],
\end{equation}
being $A = \frac{3c\Omega_{b}H_{0}^{2}}{8\pi Gm_{p}}$.

We also define $DM_{ext}$ as the difference between the $DM$ observed and its galactic contribution
\begin{equation}\label{DMext_obs}
  DM_{ext}(z) \equiv  DM_{obs}(z) - DM_{MW}\;,
\end{equation}
whereas the theoretical extragalactic dispersion measure ($DM_{ext}^{th}$) can be calculated using Eq. (\ref{DMobs})
\begin{equation}\label{DMext_th}
    DM_{ext}^{th}(z) \equiv DM_{IGM}(z) + DM_{host}(z)\;.
\end{equation} 
Thus, by using the above equations, we can compare theory and observations to constrain $f_{IGM,0}$ and $DM_{host,0}$. Following \cite{Lemos2023}, the observational data points are obtained by combining the $DM-z$ relation with $d_{L}(z)$ estimates from SNe observations.

%---------------------------------------------------------------------------------------------------------------------

%%-----------------------------------------------------------------------------
\section{Data and Methodology}\label{sec:data}
\indent

There are 19 well-localized FRBs events (for details of FRBs catalogue \footnote{https://www.herta-experiment.org/frbstats/}, see \cite{2021ascl.soft06028S}). In our analysis, we exclude the events FRB 20191228, FRB 20190614D, FRB 20190520B and FRB 20181030A due to the following reasons: FRB 20190614D \cite{FRB190614} has no measurement of spectroscopic redshift and can, in principle, be associated with two host galaxies. FRB 20190520B \cite{Ocker} has a host contribution much larger than the other FRBs, whereas FRB 20191228 \cite{FRB20191228} has the uncertainty of observed dispersion measure much larger than the others ($\sigma_{obs} = 8$ pc/cm$^{3}$); and finally, there is no SNe in the Pantheon catalogue with the redshift close to FRB 20181030A \cite{Bhardwaj2021_2} ($z = 0.0039$).

The remaining sample contains 15 FRBs with well-measured redshift, which constitutes the most up-to-date FRB data set currently available \cite{FRB180916,FRB201124,FRB190608,FRB190523_2,FRB121102,FRB190102,FRB180924,FRB181112,FRB190523_1}, and is listed in Tab. \ref{tab:FRB} with the observed dispersion measure ($DM_{obs}$), the Galaxy contribution ($DM_{MW, ISM}$) estimated from the NE2001 model \cite{Cordes2002}, and the uncertainty of $DM_{obs}$ ($\sigma_{obs}$).

The observational quantity $DM_{ext}$ (eq. \ref{DMext_obs}) can be obtained using data from Tab. \ref{tab:FRB} with its uncertainty calculated by the expression
\begin{equation}\label{uncertainty}
    \sigma_{ext}^{2} = \sigma_{obs}^{2} + \sigma_{MW}^{2}\ +\delta^{2} \;,
\end{equation}
where the average galactic uncertainty $\sigma_{MW}$ is assumed to be 10 pc/cm$^{3}$ \cite{Manchester2005} and $\delta$ stands for the $DM$ fluctuations due to the spatial variation in cosmic electron density. Such fluctuations can be treated as a probability distribution or as fixed value in the statistical analyses \cite{Wu2021,Macquart2020,Jaroszynski2019}. In this work, we will consider three different values for $\delta = 0, 100, 200, 400, 230\sqrt{z}$ pc/cm$^{3}$, in agreement with recent results presented in the literature \cite{Takahashi2021,Lemos2023}. 

We obtain the luminosity distance in Eq. (\ref{constant}) from current SNe observations, specifically the Pantheon catalogue \cite{Scolnic},  which contains 1048 SNe within the redshift range $0.01 < z < 2.3$. The distance moduli ($\mu(z)$) is given by 
\begin{equation} \label{mz}
    \mu(z) = m_{B} - M_{B} = 5\log_{10}\left[ \frac{d_{L}(z)}{1\mbox{Mpc}}\right] + 25 \;,
\end{equation}
where $m_{B}$ and $M_{B}$ are the apparent magnitude of SNe and the absolute peak magnitude, respectively. In our analysis we fix $M_{B} = -19.214 \pm 0.037$ mag \cite{Riess2019} or, equivalently, $H_{0} = 74.03 \pm 1.4$ kms$^{-1}$Mpc$^{-1}$. %, and $\Omega_{b}h^{2} =  0.02235 \pm 0.00037$ from a recent Big Bang Nucleosynthesis (BBN) study \cite{Cooke2018}. 
To obtain estimates of $d_{L}(z)$ at the same redshift of the FRBs, we perform a Gaussian Process (GP) reconstruction of the Pantheon data, using  GaPP python library (for details of GaPP\footnote{https://github.com/astrobengaly/GaPP}, see \cite{GaPP}). There are two free parameters ($f_{IGM,0}$, $DM_{host,0}$) in Eq. (\ref{constant}), which will be constrained from the Monte Carlo Markov Chain (MCMC) analysis using the \textit{emcee} package \cite{Foreman-Mackey2013}. The results of our observational data analysis for $\delta = 0, 100, 200, 400, 230\sqrt{z}$ pc/cm$^{3}$ are displayed in Table \ref{tab:observations}.%  (see also reference \cite{Lemos2023} for more details).

\begin{table}
%\vspace{0.3cm}
\centering
\caption{Estimates of the $f_{IGM}$ and $DM_{host,0}$ from current observational data.}
\begin{tabular}{ c  c  c}
\hline
$\delta$ [pc/cm$^{3}$] & $f_{IGM,0}$ & $DM_{host,0}$ [pc/cm$^{3}$]\\
\hline
0   & 0.77 $\pm$ 0.01 & 158.8 $\pm$ 5.3 \\
100  & 0.76 $\pm$ 0.11 & 158.0 $\pm$ 50.0 \\
200 & 0.74 $\pm$ 0.16 & 152.0 $\pm$ 65.0 \\
400 & 0.66 $\pm$ 0.17 & 142.0 $ \pm$ 70.0 \\
230$\sqrt{z}$ & 0.81 $\pm$ 0.12 & 133.0 $\pm$ 30.0\\
\hline
%\multicolumn{6}{p{2.5cm}}{\,} 
\end{tabular}
\label{tab:observations}
\end{table}

%%-----------------------------------------------------------------------------

\begin{figure}
\includegraphics[width=0.47\textwidth]{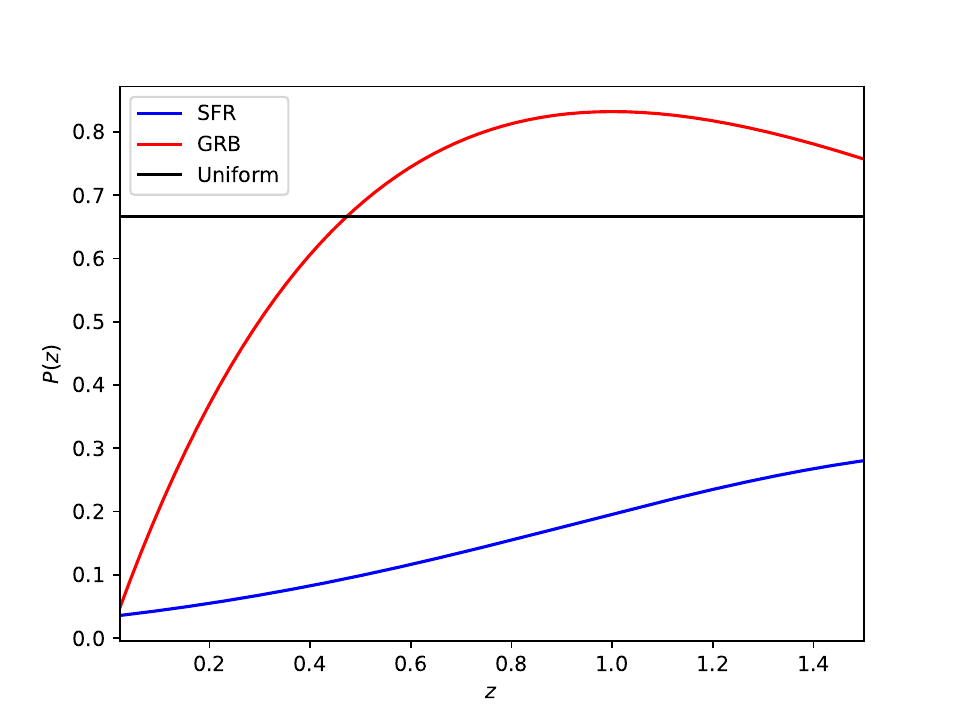}
%\vspace{-0.5cm}
\caption{The normalized redshift distributions for FRBs.}
\label{fig:PDF}
%colour_magnitude_diagram
\end{figure}

\begin{figure*}
\begin{center}
\includegraphics[width=0.523\textwidth]{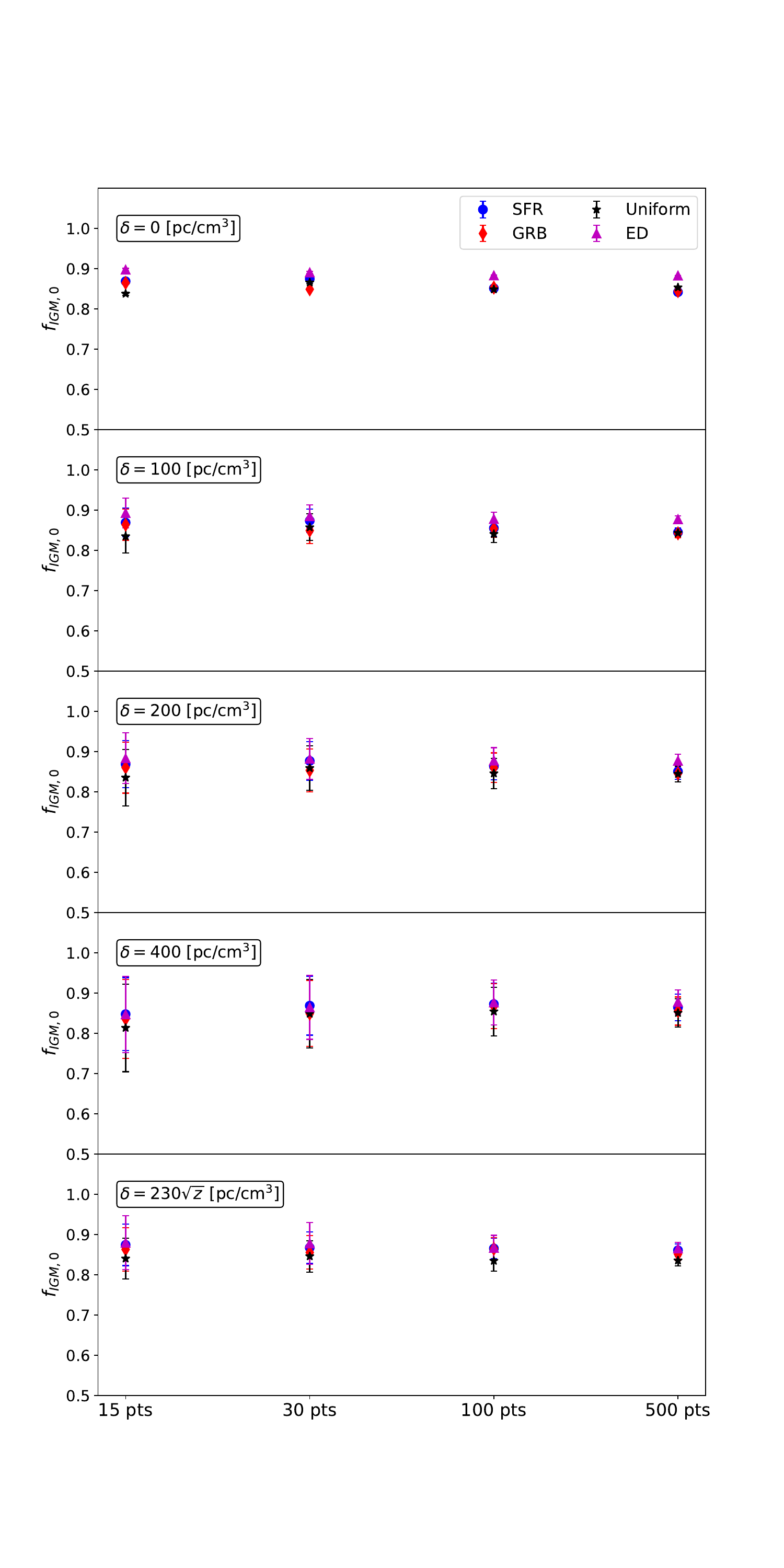}
\hspace{-1.0cm}
\includegraphics[width=0.523\textwidth]{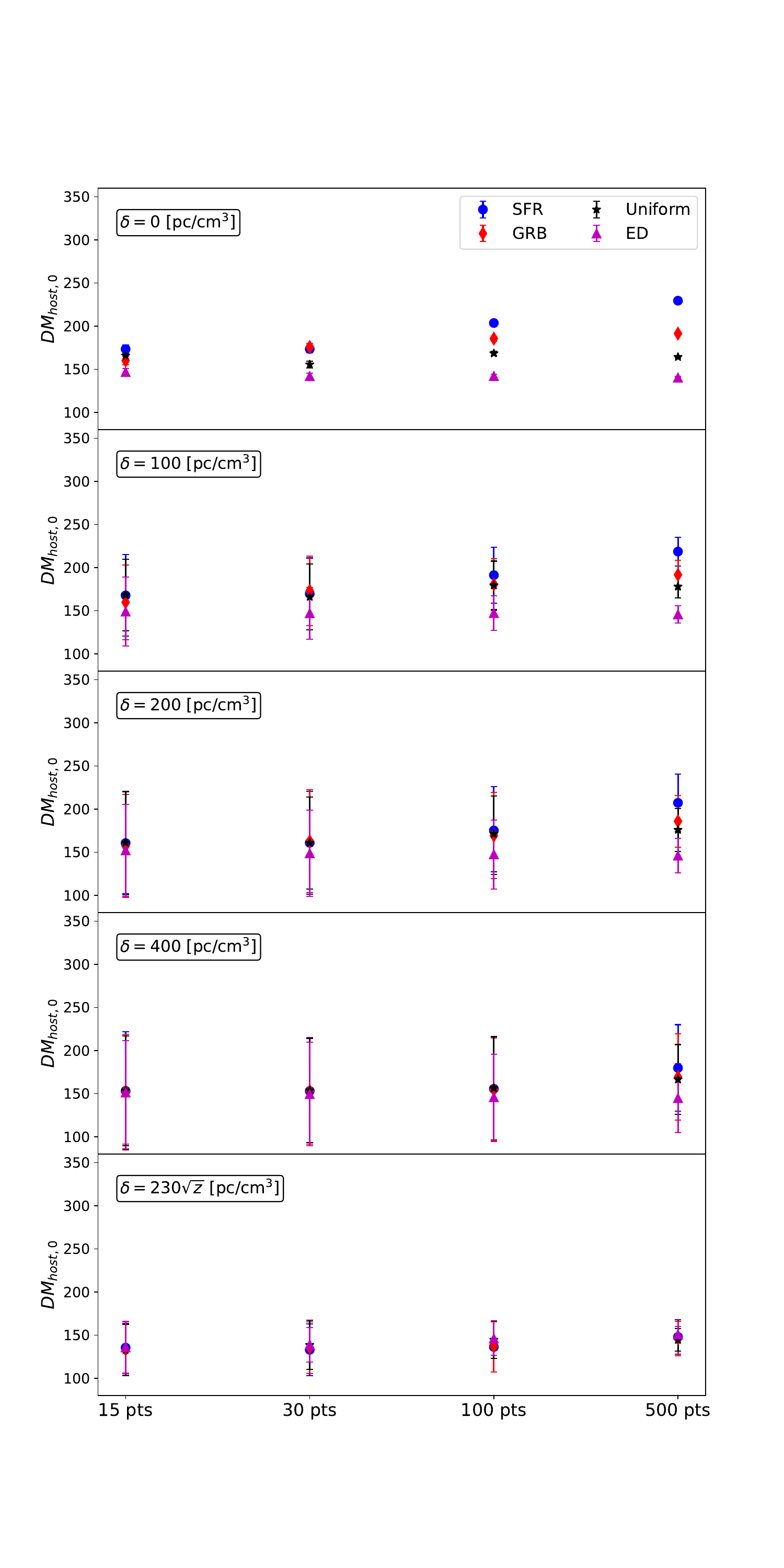}
\caption{The results of our simulations for $f_{IGM,0}$ and $DM_{host,0}$. The data points represent the average values of these parameters for each distribution model discussed in the text, considering different sizes of sample and values of $DM$ fluctuations. } 
\label{fig:results}
%colour_magnitude_diagram
\end{center}
\end{figure*}

\section{Simulations}\label{sec:simulations}

To study the cosmological impact of a larger sample of FRBs than the one currently available, we perform a Monte Carlo simulation to generate random points of $DM_{ext}$.  For the MC simulation method, we need a redshift distribution of FRBs to generate the points, but the distribution of these bursts is still uncertain because we do not know the progenitor of these events, and for this reason many models for distribution of FRBs have been assumed. In reference \cite{Qiang2021}, the authors studied the effects of nine different redshift distribution of FRBs to constrain cosmological parameters and found that three of them present strong constraining power. Thus, we will consider these three distributions, namely: %gamma-ray bursts (GRBs), star formation rate (SFR) and uniform case (Uniform). 

\begin{itemize}
    
    \item {\it{Gamma-Ray Bursts:}} Several studies assume the gamma-ray bursts distribution for FRBs due to the similarity between these two events \cite{GRB}. The density function is written as %(termed 'PGRB'). 
\begin{equation}
    P_{\rm{GRB}}(z) \propto z  \exp{(-z)}.
\end{equation}

    \item {\it{Star Formation Rate:}} The star formation rate distribution was proposed by \cite{Madau} (see also reference \cite{Bhattacharya} for the first proposal of redshift distribution for FRBs). The spatial distribution of FRBs is expected to closely trace the cosmic one for young stellar FRB progenitors. The cosmic SFR function can be written as
\begin{equation}
    \psi(z) = 0.015 \frac{(1+z)^{2.7}}{1+[(1+z)/2.9]^{5.6}} . 
\end{equation}

    \item {\it{Uniform:}} The uniform distribution %(termed 'PUniform')
assumes that the FRBs distribution is constant and its density function is given by
\begin{equation}
    P_{\rm{Uniform}} = \frac{1}{z_{max} - z_{min}}. 
\end{equation}

\end{itemize}
For completeness, we also consider an additional distribution, where the FRBs redshifts are picked at equidistant points (ED) between $z_{min}$ and $z_{max}$.

%In Figure \ref{fig:PDF} we present the three redshift distribution models for FRBs. As we are interested in comparing our results with the observational data, we will simulate the data points in the same range of FRBs redshifts presented in Table (\ref{tab:FRB}), {{$0.022 \leq z \leq 0.66$}}. %The choice of the redshift range was due to the lower SNe and upper FRBs redshifts. 

In Figure \ref{fig:PDF} we present the three redshift distribution models for FRBs. Since for $z > 1.5$, the GP reconstruction of the Pantheon data overestimates the uncertainty values (given the small number of points in such interval), we will simulate data points in the  {{$0.022 \leq z \leq 1.5$}} interval.

The steps of our simulations are the following: 
\begin{enumerate}
    \item We generate random points using the redshift distribution models described above in the redshift range \textbf{$[0.022, 1.5]$}. We consider samples with N = 15, 30, 100 and 500 points. 

    \item We calculate the fiducial $DM_{ext}$ ($DM_{ext}^{fid}$) using Eq. (\ref{DMext_th}), where $DM_{IGM}$ is given by Eq. (\ref{IGM}). We adopt the mean values of baryon fraction and host contribution as reported in \cite{Lemos2023} for the constant case, i.e., $f_{IGM,0} = 0.764$ and $DM_{host,0} = 158.1$ pc/cm$^{3}$. In our simulations, we also adopt the values of $H_{0} = 74.03 \pm 1.4$ kms$^{-1}$Mpc$^{-1}$ \cite{Riess2019}, $\Omega_{m} = 0.3153 $ \cite{Planck} and $\Omega_{b}h^{2} =  0.02235 \pm 0.00037$ \cite{Cooke2018}.

    \item We calculate the uncertainty of $DM_{ext}$ simulated ($\sigma_{ext}^{sim}$). The $DM_{IGM}$ and $DM_{host,0}$ uncertainties are not well constrained, so we calculate $\sigma_{ext}^{sim}$ performing a regression of observational data of relative error. As long as the relative error decreases with $z$ and cannot be negative, we consider the relative error described by an hyperbolic function which is $\eta = \sigma_{ext}^{obs}/DM_{ext}^{obs} = A/z $, where $A$  hyperbolic regression free parameter. 

    \item Finally, we calculate the simulated $DM_{ext}$ by assuming a normal distribution, given by $DM_{ext}^{sim} (z) = \mathcal{N}(DM_{ext}^{fid},sd)$. Here, $sd$ represents the standard deviation of the Gaussian Distribution, which is obtained from the average distance between the observed and fiducial points.
    
\end{enumerate}
We perform the steps above 50 times for each
sample size of the distribution models, which is enough to obtain convergence (see Appendix A). In each simulation, we calculate the free parameters while considering different values of $DM$ fluctuations $\delta = 0, 100, 200, 400, 230\sqrt{z}$ pc/cm$^{3}$. Regarding the $DM_{host,0}$, we assume in our MCMC analysis a Gaussian prior for this parameter, with the mean
value and standard deviation being the best-fit values shown in Table  \ref{tab:observations}. Subsequently, we calculate the average of each ensemble of 50 simulations.

\begin{table*}
%\begin{sidewaystable}
\centering
%\begin{center}
\scalebox{0.83}{
\renewcommand{\arraystretch}{1.3}
\begin{tabular}{|c|c|c|c|c|c|c|c|c|}
%\hline
\hline
 $N$ & $f_{IGM,0}$ & $DM_{host,0}$  & $f_{IGM,0}$ & $DM_{host,0}$  & $f_{IGM,0}$ & $DM_{host,0}$  & $f_{IGM,0}$ & $DM_{host,0}$  \\
&  &   [pc/cm$^{3}$] &  & [pc/cm$^{3}$] &  & [pc/cm$^{3}$] &  & [pc/cm$^{3}$] \\
\hline
& \multicolumn{2}{|c|}{SFR} & \multicolumn{2}{|c|}{GRB}  & \multicolumn{2}{|c|}{Uniform} & \multicolumn{2}{|c|}{ED}  \\ 
%& SFR &  &  & GRB & Uniform & & ED &  \\ 
\hline
\hline
   &  & & & $\delta= 0$  pc/cm$^{3}$ &  & &  &  \\ 
   \hline
   15   & 0.8686 $\pm$ 0.0032 & 173.43 $\pm$ 4.82 & 0.8645 $\pm$ 0.0034 & 160.05 $\pm$ 4.66 & 0.8377 $\pm$ 0.0037 & 165.83 $\pm$ 4.47 & 0.8975 $\pm$ 0.0032 & 146.79 $\pm$ 4.30  \\
   30   & 0.8743 $\pm$ 0.0026 & 173.55 $\pm$ 4.34 & 0.8487 $\pm$ 0.0029 & 175.80 $\pm$ 4.15 & 0.8656 $\pm$ 0.0030 & 155.59 $\pm$ 3.89 & 0.8905 $\pm$ 0.0025 & 141.92 $\pm$ 3.60  \\
   100  & 0.8512 $\pm$ 0.0018 & 203.64 $\pm$ 3.21 & 0.8529 $\pm$ 0.0019 & 185.53 $\pm$ 2.97 & 0.8487 $\pm$ 0.0019 & 168.44 $\pm$ 2.66 & 0.8832 $\pm$ 0.0015 & 142.08 $\pm$ 2.40  \\
   500  & 0.8417 $\pm$ 0.0009 & 229.52 $\pm$ 1.69 & 0.8442 $\pm$ 0.0009 & 191.39 $\pm$ 1.52 & 0.8534 $\pm$ 0.0009 & 164.28 $\pm$ 1.31 & 0.8827 $\pm$ 0.0007 & 140.31 $\pm$ 1.20  \\
   \hline
   &  & & & $\delta= 100$  pc/cm$^{3}$ &  & &  &  \\ 
   \hline
   15   & 0.8689 $\pm$ 0.0357 & 167.76 $\pm$ 47.22 & 0.8638 $\pm$ 0.0385 & 159.74 $\pm$ 43.45 & 0.8350 $\pm$ 0.0418 & 168.36 $\pm$ 41.54 & 0.8927 $\pm$ 0.0372 & 149.08 $\pm$ 40.00  \\
   30   & 0.8737 $\pm$ 0.0294 & 169.66 $\pm$ 41.55 & 0.8491 $\pm$ 0.0320 & 173.18 $\pm$ 40.22 & 0.8573 $\pm$ 0.0332 & 166.10 $\pm$ 38.03 & 0.8847 $\pm$ 0.0288 & 147.08 $\pm$ 30.00  \\
   100  & 0.8551 $\pm$ 0.0201 & 191.28 $\pm$ 32.25 & 0.8537 $\pm$ 0.0214 & 180.48 $\pm$ 30.00 & 0.8406 $\pm$ 0.0209 & 179.40 $\pm$ 28.19 & 0.8779 $\pm$ 0.0170 & 147.34 $\pm$ 20.00  \\
   500  & 0.8447 $\pm$ 0.0108 & 218.62 $\pm$ 16.65 & 0.8423 $\pm$ 0.0110 & 191.68 $\pm$ 16.79 & 0.8434 $\pm$ 0.0100 & 177.98 $\pm$ 13.21 & 0.8774 $\pm$ 0.0080 & 145.78 $\pm$ 10.02  \\
   \hline
   &  & & & $\delta= 200$  pc/cm$^{3}$ &  & &  &  \\ 
   \hline
   15   & 0.8685 $\pm$ 0.0586 & 160.56 $\pm$ 60.00 & 0.8603 $\pm$ 0.0635 & 157.12 $\pm$ 59.82 & 0.8353 $\pm$ 0.0699 & 161.04 $\pm$ 59.26 & 0.8833 $\pm$ 0.0631 & 152.12 $\pm$ 53.61  \\
   30   & 0.8768 $\pm$ 0.0479 & 160.88 $\pm$ 59.45 & 0.8533 $\pm$ 0.0528 & 162.64 $\pm$ 59.63 & 0.8590 $\pm$ 0.0555 & 160.78 $\pm$ 53.40 & 0.8821 $\pm$ 0.0507 & 148.82 $\pm$ 50.00  \\
   100  & 0.8634 $\pm$ 0.0340 & 175.20 $\pm$ 50.87 & 0.8601 $\pm$ 0.0367 & 169.30 $\pm$ 50.00 & 0.8459 $\pm$ 0.0377 & 171.20 $\pm$ 44.07 & 0.8777 $\pm$ 0.0320 & 147.52 $\pm$ 40.00  \\
   500  & 0.8505 $\pm$ 0.0199 & 207.24 $\pm$ 33.32 & 0.8456 $\pm$ 0.0207 & 185.92 $\pm$ 30.00 & 0.8445 $\pm$ 0.0195 & 175.98 $\pm$ 24.90 & 0.8772 $\pm$ 0.0160 & 146.04 $\pm$ 20.00  \\
    \hline
   &  & & & $\delta = 400$  pc/cm$^{3}$ &  & &  &  \\ 
   \hline
   15   & 0.8475 $\pm$ 0.0908 & 153.26 $\pm$ 68.50 & 0.8356 $\pm$ 0.0984 & 152.44 $\pm$ 66.38 & 0.8135 $\pm$ 0.1090 & 153.32 $\pm$ 63.58 & 0.8470 $\pm$ 0.0951 & 151.30 $\pm$ 60.00  \\
   30   & 0.8686 $\pm$ 0.0731 & 153.12 $\pm$ 61.71 & 0.8489 $\pm$ 0.0818 & 153.06 $\pm$ 61.29 & 0.8488 $\pm$ 0.0851 & 153.84 $\pm$ 60.32 & 0.8645 $\pm$ 0.0794 & 149.54 $\pm$ 60.00  \\
   100  & 0.8725 $\pm$ 0.0515 & 155.64 $\pm$ 60.00 & 0.8674 $\pm$ 0.0558 & 154.42 $\pm$ 60.00 & 0.8542 $\pm$ 0.0601 & 156.32 $\pm$ 60.00 & 0.8766 $\pm$ 0.0553 & 145.90 $\pm$ 50.00  \\
   500  & 0.8645 $\pm$ 0.0327 & 179.80 $\pm$ 50.22 & 0.8552 $\pm$ 0.0351 & 169.30 $\pm$ 50.00 & 0.8507 $\pm$ 0.0352 & 166.50 $\pm$ 40.45 & 0.8779 $\pm$ 0.0300 & 144.80 $\pm$ 40.00  \\
   \hline
   &  & & & $\delta = 230\sqrt{z}$  pc/cm$^{3}$ &  & &  &  \\ 
   \hline
   15   & 0.8743 $\pm$ 0.0516 & 135.50 $\pm$ 30.00 & 0.8629 $\pm$ 0.0538 & 133.36 $\pm$ 30.00 & 0.8403 $\pm$ 0.0502 & 132.90 $\pm$ 29.83 & 0.8797 $\pm$ 0.0666 & 135.78 $\pm$ 30.00  \\
   30   & 0.8674 $\pm$ 0.0391 & 133.00 $\pm$ 30.00 & 0.8554 $\pm$ 0.0416 & 135.38 $\pm$ 30.00 & 0.8460 $\pm$ 0.0390 & 138.60 $\pm$ 28.46 & 0.8778 $\pm$ 0.0522 & 138.92 $\pm$ 20.00  \\
   100  & 0.8654 $\pm$ 0.0257 & 136.32 $\pm$ 29.32 & 0.8646 $\pm$ 0.0272 & 137.00 $\pm$ 29.66 & 0.8347 $\pm$ 0.0253 & 144.66 $\pm$ 21.39 & 0.8670 $\pm$ 0.0310 & 146.08 $\pm$ 19.58  \\
   500  & 0.8608 $\pm$ 0.0156 & 147.90 $\pm$ 20.00 & 0.8511 $\pm$ 0.0161 & 146.32 $\pm$ 20.00 & 0.8354 $\pm$ 0.0135 & 144.52 $\pm$ 13.03 & 0.8651 $\pm$ 0.0150 & 150.75 $\pm$ 9.40   \\
\hline
\end{tabular}
}
\caption{The results of our simulations for $f_{IGM,0}$ and $DM_{host,0}$ considering the distribution models discussed in the text. }%, $N$ = 15, 30, 100 and 500 points and $\delta = 0, 10, 100$ pc/cm$^{3}$.}
\label{tab:results}
%\end{center}
%\end{sidewaystable} 
\end{table*}

%%-----------------------------------------------------------------------------

\section{Results}\label{sec:results}

The results of our simulations are displayed in Fig. \ref{fig:results} and Table \ref{tab:results}. In Figure \ref{fig:results}, we present the 1$\sigma$ error bars  for the free parameters $f_{IGM,0}$ and $DM_{host,0}$, considering different redshift distributions and values of $\delta = 0, 100, 200, 400, 230\sqrt{z}$ pc/cm$^{3}$. Table \ref{tab:results} shows the numerical values obtained separately for all distributions and different numbers of points in each realization ($N = 15, 30, 100, 500$).

%For all distributions {{(except for the ED distribution)}} the constraints on $f_{IGM,0}$ and $DM_{host,0}$ are compatible within $2\sigma$. Comparing the results of simulations for $N = 15$ with the results for the current observational data (which also comprises 15 points), we find that:  (i) for $\delta = 0$  pc/cm$^{3}$, GRB and SFR distributions show a good agreement for $f_{IGM,0}$ within $1\sigma$ while the results from the Uniform distribution are compatible only at $2\sigma$; (ii) for $DM_{host,0}$, differently from the SFR distribution, GRB, Uniform and ED distributions agree at $2\sigma$; (iii) the agreement between the different redshift distributions becomes stronger as the value of the DM fluctuation increases. In particular, for $\delta = 100$  pc/cm$^{3}$, the results from the GRB, SFR and Uniform distributions agree at $1\sigma$ for both parameters $f_{IGM,0}$ and $DM_{host,0}$.

For all distributions {{(except for the sample $N =15$)}} the constraints on $f_{IGM,0}$ and $DM_{host,0}$ are compatible within $2\sigma$. Comparing the results of simulations for $N = 15$ with the results for the current observational data (which also comprises 15 points), we find that:  (i) for $\delta = 0$  pc/cm$^{3}$, all distributions are not in agreement for $f_{IGM,0}$ within $2\sigma$; (ii) for $DM_{host,0}$, differently from the SFR distribution, GRB, Uniform and ED distributions agree at $2\sigma$; (iii) for the other values of the $DM$ fluctuation, the results from the redshift distributions are in agreement within $1\sigma$ for both parameters $f_{IGM,0}$ and $DM_{host,0}$.

Finally, it is worth mentioning that the errors on the $f_{IGM,0}$ and $DM_{host,0}$ parameters depend on the number of points and the $DM$ fluctuations. Our results show that such errors are smaller for a given value of the $DM$ fluctuations as larger number of points is considered. On the other hand, the errors increase for results with the same number of points $N$ and higher values of $\delta$. Therefore, these results show that larger data samples, as expected by the next generations of surveys, play a crucial role in this kind of analysis, along with a better understanding of the $DM$ fluctuations parameter.

%Finally, it is worth mentioning that the errors are dependent on the number of points as well as the $DM$ fluctuations. Moreover, for the results with the same $DM$ fluctuations, but with larger number of points, the errors are smaller. On the other hand, for the results with the same $N$ and higher values of $\delta$, the errors are greater. These results show that larger samples of data, expected by the next generations of surveys, play a crucial role in the analysis, along with a better understanding of the $DM$ fluctuations parameter.

%Finally, it is worth mentioning that for each distribution the errors bars obtained for the largest sample ($N = 500$) and largest DM fluctuations ($\delta = 100$ pc/cm$^{3}$) are bigger than those derived with the smallest sample ($N = 15$) and the smallest fluctuations ($\delta = 0$). These results clearly show that the $DM$ fluctuations play a crucial role in our analysis, and a better understanding of this parameter is essential in obtaining more precise estimates of cosmological parameters from FRB data.

%%-----------------------------------------------------------------------------
\section{Conclusions}\label{sec:conclusions}

FRB observations have demonstrated a great potential to constrain cosmological parameters and test aspects of fundamental physics. In this context, although some of their astrophysical characteristics are still under debate, the growing significance of these transient events in cosmology is becoming apparent. Therefore, it is important to investigate the constraining power of upcoming FRB observations on physical and cosmological parameters to better understand their potential and limitations.

In this work, we investigated the impact of the $DM$ fluctuations and the number of FRBs observations to constrain the parameters $f_{IGM,0}$ and $DM_{host,0}$ from simulated data considering distinct probability distributions for the sources. Firstly, we performed a statistical analysis with 15 observational data points following the model-independent method presented in \cite{Lemos2023}. Our sample was defined from an original sample of 20 data points, where we removed five sources for different reasons, e.g. discrepant values for the uncertainties or redshift incompatibility with the SNe catalogue. Secondly, we generated data sets  from Monte Carlo simulations considering four redshift distributions, namely Gamma-ray Bursts, Star Formation Rate, Uniform and Equidistant distributions. The number of points in the analyses varied from $ N = 15, 30, 100, 500$, as expected from upcoming projects, whereas the $DM$ fluctuations assumed values of $\delta = 0, 100, 200, 400, 230\sqrt{z}$ pc/cm$^{3}$. 

%The results showed an agreement within 2$\sigma$ between the GRB, SFR and Uniform distributions, regardless of the values of $\delta$. In particular, our analysis highlighted the crucial role of  $DM$ fluctuations in the results, which reinforces the need for more investigations into this quantity. As an example, for $N = 100$, as expected by the ASKAP/CRACO per year \cite{James2022}, we found that the expected relative error for $f_{IGM,0}$ varies from $\sim 0.6\%$ ($\delta = 0$ pc/cm$^{3}$) to $8\%$ ($\delta = 100$ pc/cm$^{3}$) and from $\sim 2\%$ ($\delta = 0$ pc/cm$^{3}$) to $20\%$ ($\delta = 100$ pc/cm$^{3}$) for $DM_{host,0}$ (see Table~\ref{tab:results}).

The results showed an agreement within 2$\sigma$ between the GRB, SFR, Uniform and ED distributions, regardless of the values of $\delta$. In particular, our analysis highlighted the crucial role of  $DM$ fluctuations in the results, which reinforces the need for more investigations into this quantity. As an example, for $N = 100$, as expected by the ASKAP/CRACO per year \cite{James2022}, we found that the expected relative error for $f_{IGM,0}$ varies from $\sim 0.2\%$ ($\delta = 0$ pc/cm$^{3}$) to $6\%$ ($\delta = 400$ pc/cm$^{3}$) and from $\sim 2\%$ ($\delta = 0$ pc/cm$^{3}$) to $60\%$ ($\delta = 400$ pc/cm$^{3}$) for $DM_{host,0}$ (see Table~\ref{tab:results}).

Finally, we would like to emphasize that the method and simulated data generated in our analysis can be used to forecast model-independent constraints on astrophysical and cosmological parameters, as reported in this paper, and investigate expected limits on the physical parameters of fundamental theories. Some applications are in progress and will appear in a future communication.

%With the present discussion we aimed to discuss the impact of the next generations of FRBs surveys on some astrophysical and cosmological parameters. We expect that new observations can confirm this predictions and assure the FRB as a tool for precise cosmology.

%%-----------------------------------------------------------------------------

\section*{Acknowledgements}

TL thanks the financial support from the Coordena\c{c}\~ao de perfei\c{c}oamento de Pessoal de N\'{\i}vel Superior (CAPES). JSA is supported by Conselho Nacional de Desenvolvimento Cient\'{\i}fico e Tecnol\'ogico (CNPq 310790/2014-0) and Funda\c{c}\~ao de Amparo \`a Pesquisa do Estado do Rio de Janeiro (FAPERJ) grant 259610 (2021). This work was developed thanks to the High-Performance
Computing Center at the National Observatory (CPDON).

%%-----------------------------------------------------------------------------
%\section*{Data Availability}

%The inclusion of a Data Availability Statement is a requirement for articles published in MNRAS. Data Availability Statements provide a standardised format for readers to understand the availability of data underlying the research results described in the article. The statement may refer to original data generated in the course of the study or to third-party data analysed in the article. The statement should describe and provide means of access, where possible, by linking to the data or providing the required accession numbers for the relevant databases or DOIs.


%%%%%%%%%%%%%%%%%%%%%%%%%%%%%%%%%%%%%%%%%%%%%%%%%%

%%%%%%%%%%%%%%%%% APPENDICES %%%%%%%%%%%%%%%%%%%%%
\onecolumn
\appendix
\section{\centering Supplementary Material} \label{Appendix}

%\appendix{}
%\section{Supplementary Figures}\label{Appendix}

\begin{figure*}[!h]
\begin{center}
\resizebox{200pt}{130pt}{\includegraphics{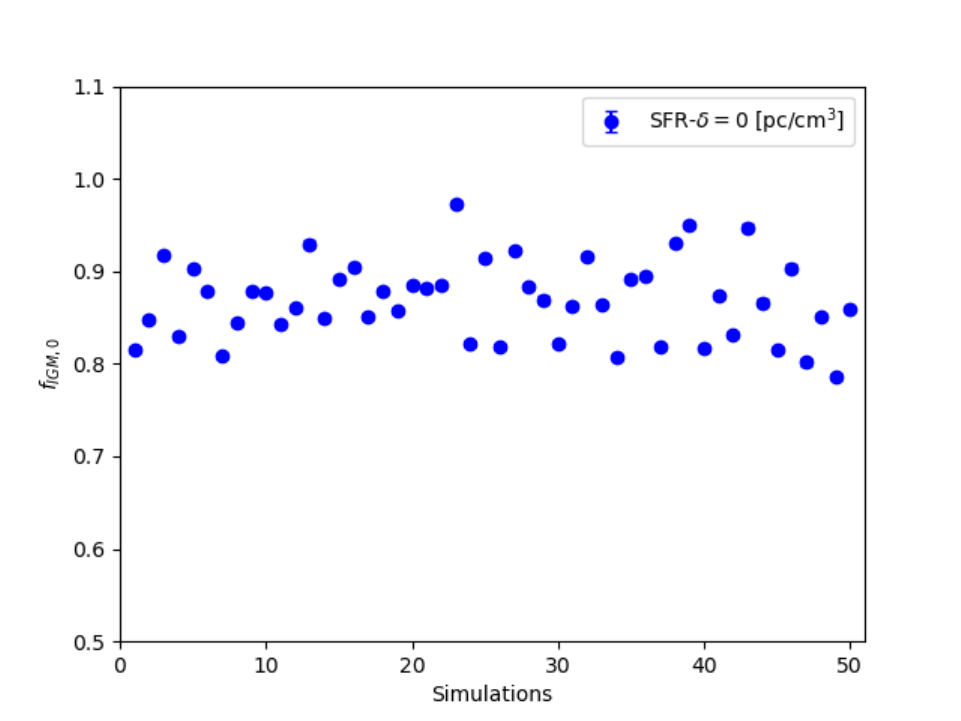}}
\hspace{1mm}\resizebox{200pt}{130pt}{\includegraphics{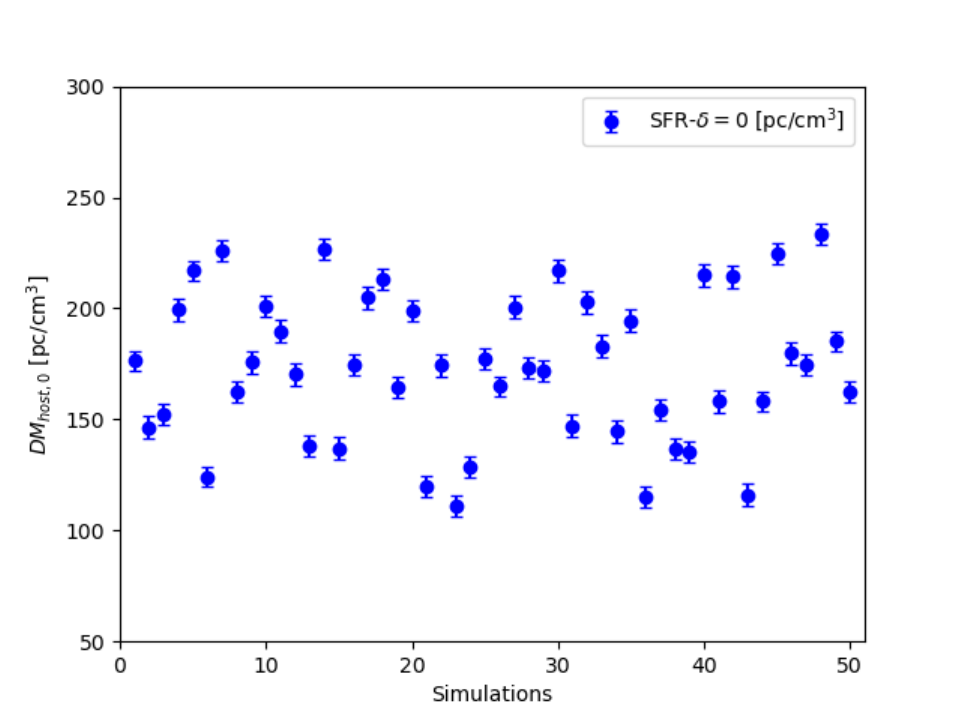}}\\
\hspace{1mm}\resizebox{200pt}{130pt}{\includegraphics{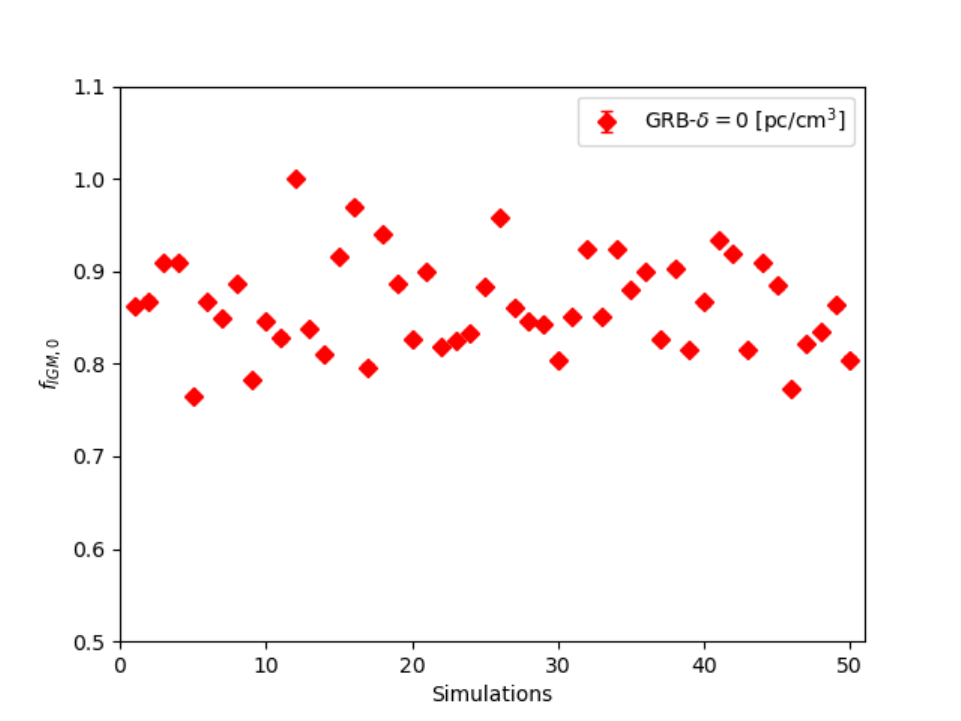}}
\resizebox{200pt}{130pt}{\includegraphics{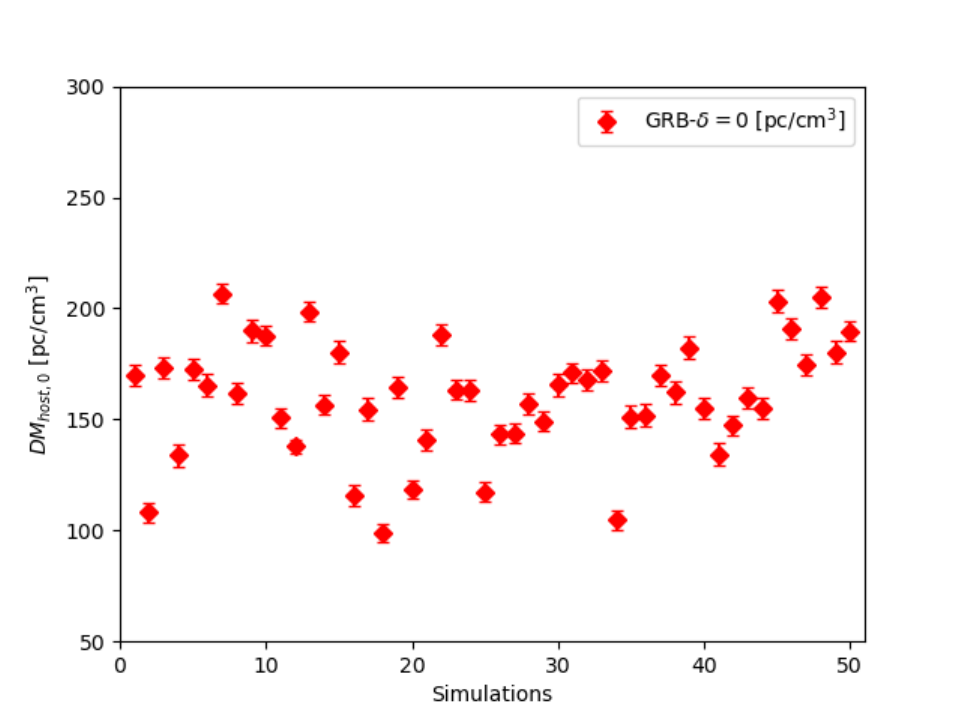}} 
\resizebox{200pt}{130pt}{\includegraphics{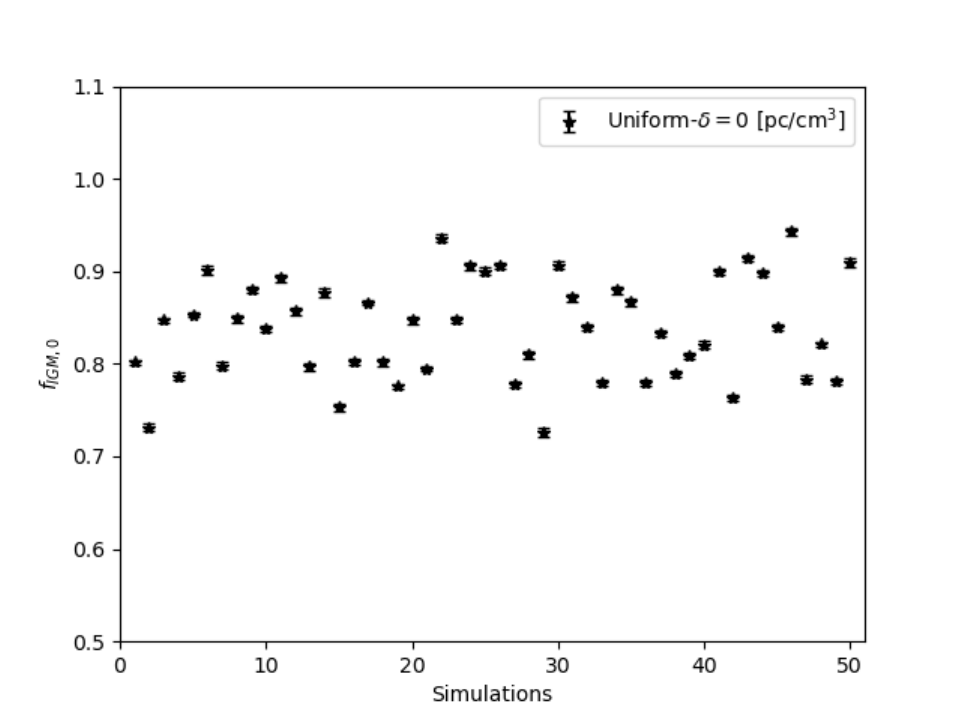}}
\hspace{1mm}\resizebox{200pt}{130pt}{\includegraphics{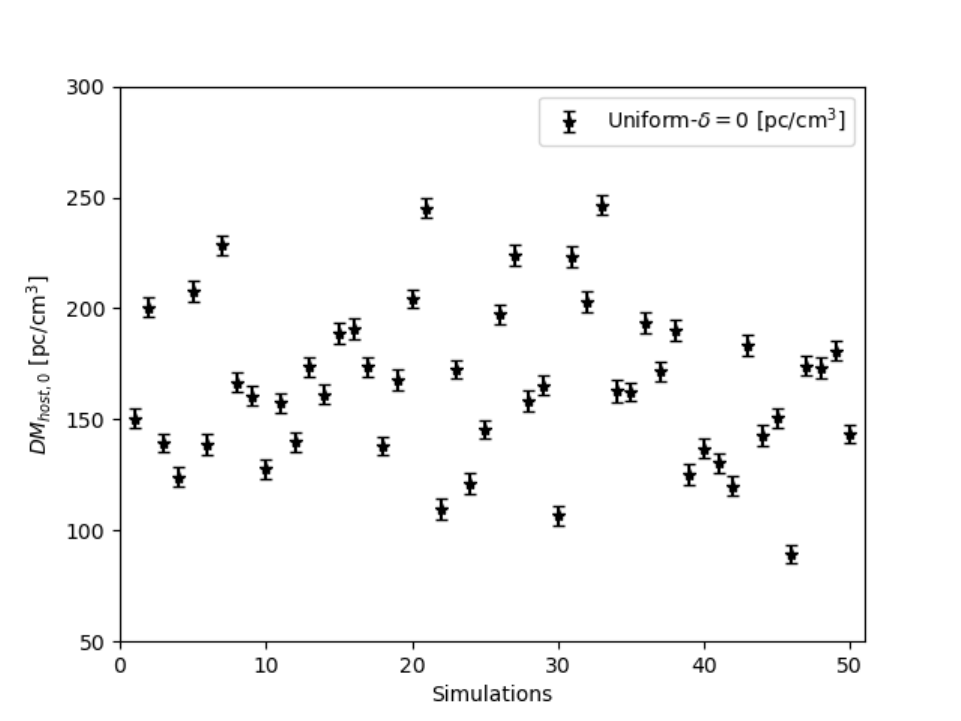}}
\resizebox{200pt}{130pt}{\includegraphics{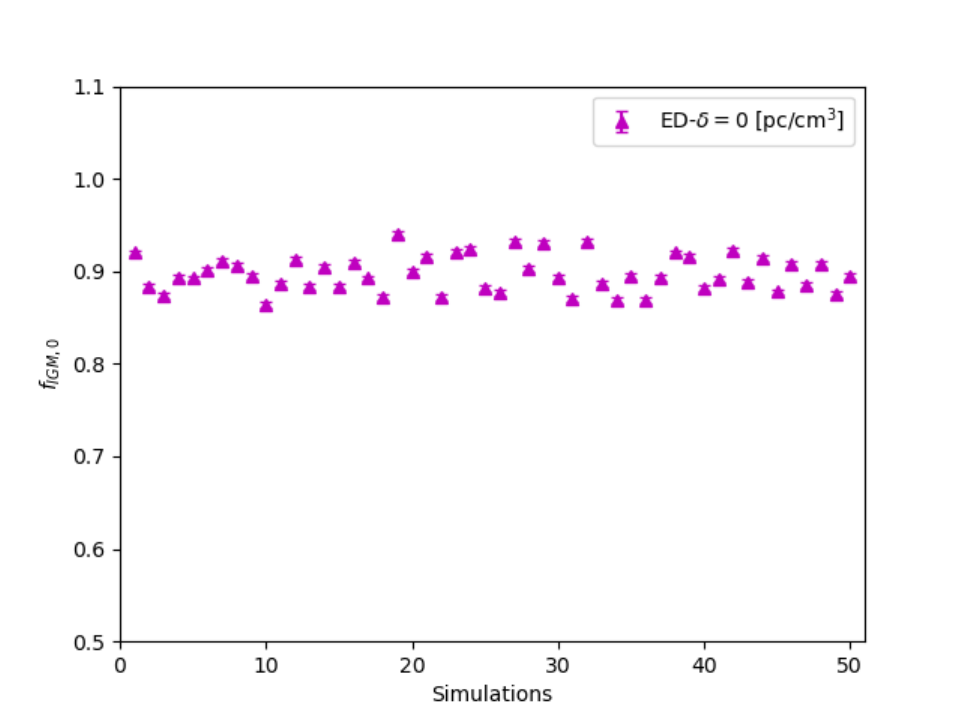}}
\hspace{1mm}\resizebox{200pt}{130pt}{\includegraphics{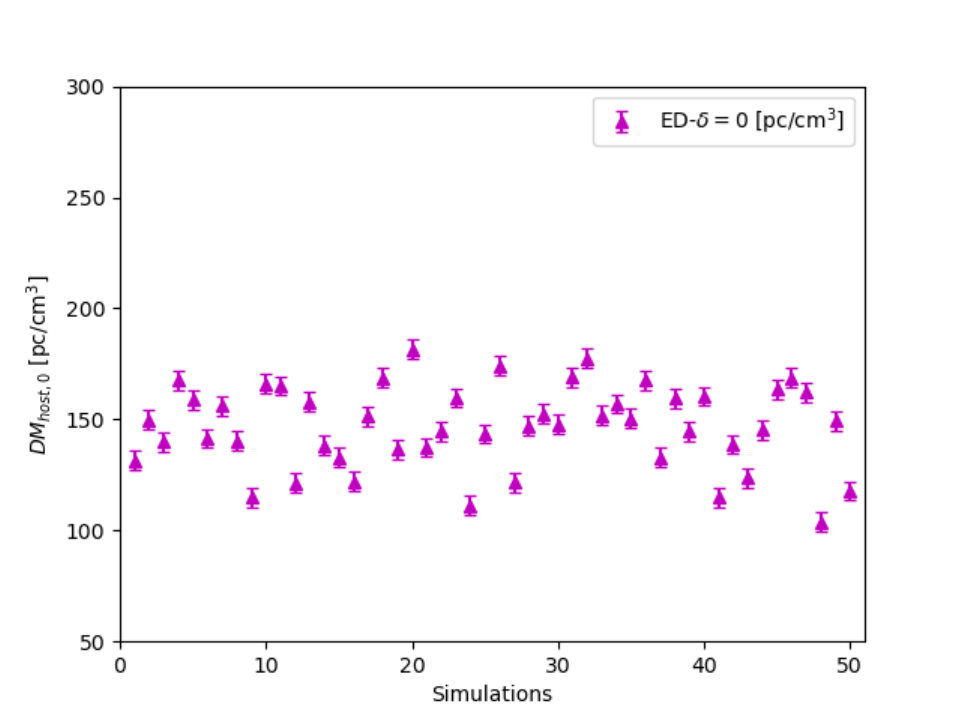}}\\
\end{center}
\caption{The best-fit of the 50 simulations of SFR, GRB, Uniform and ED considering $N = 15$ and $\delta = 0$ pc/cm$^{3}$ for both parameters $f_{IGM,0}$ and $DM_{host,0}$.}  
\label{lowexp}
\end{figure*}

\begin{figure*}[!h]
\begin{center}
\resizebox{200pt}{130pt}{\includegraphics{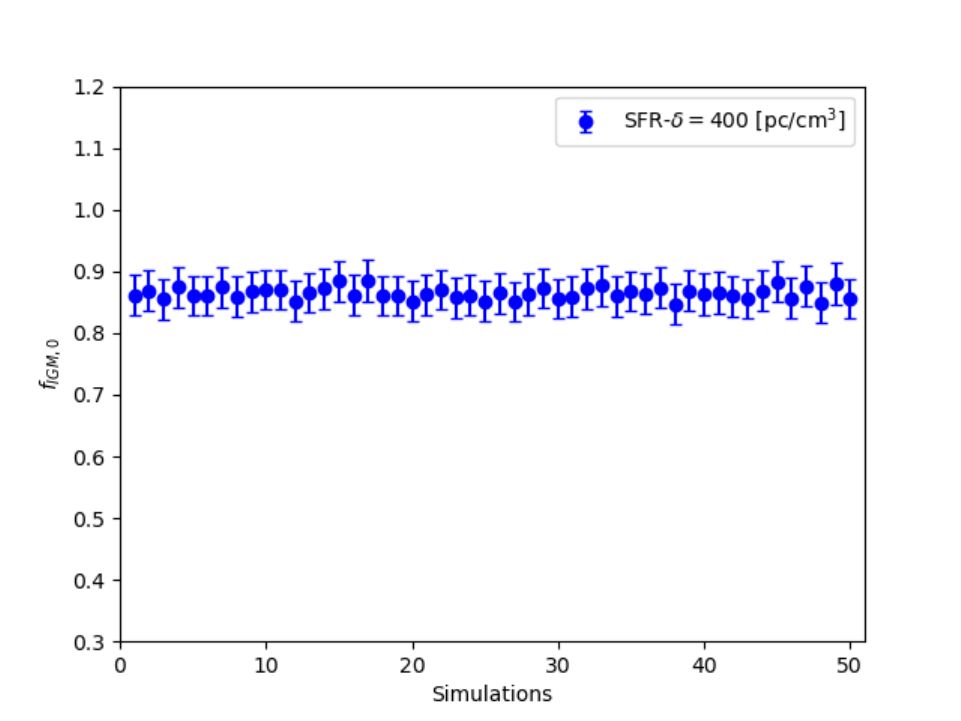}}
\hspace{1mm}\resizebox{200pt}{130pt}{\includegraphics{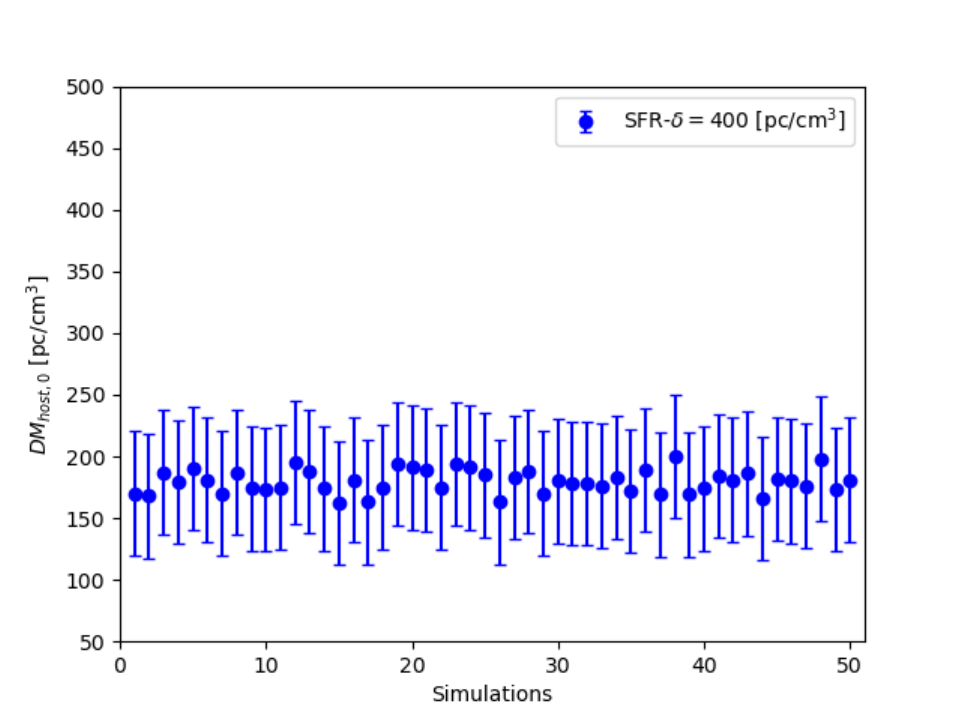}}\\
\hspace{1mm}\resizebox{200pt}{130pt}{\includegraphics{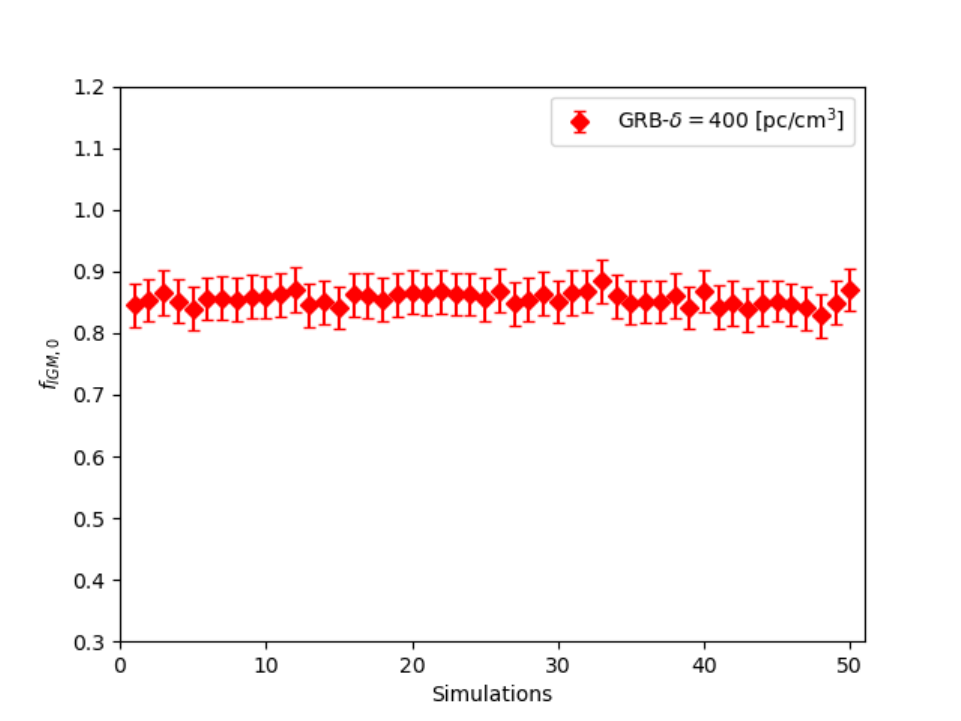}}
\resizebox{200pt}{130pt}{\includegraphics{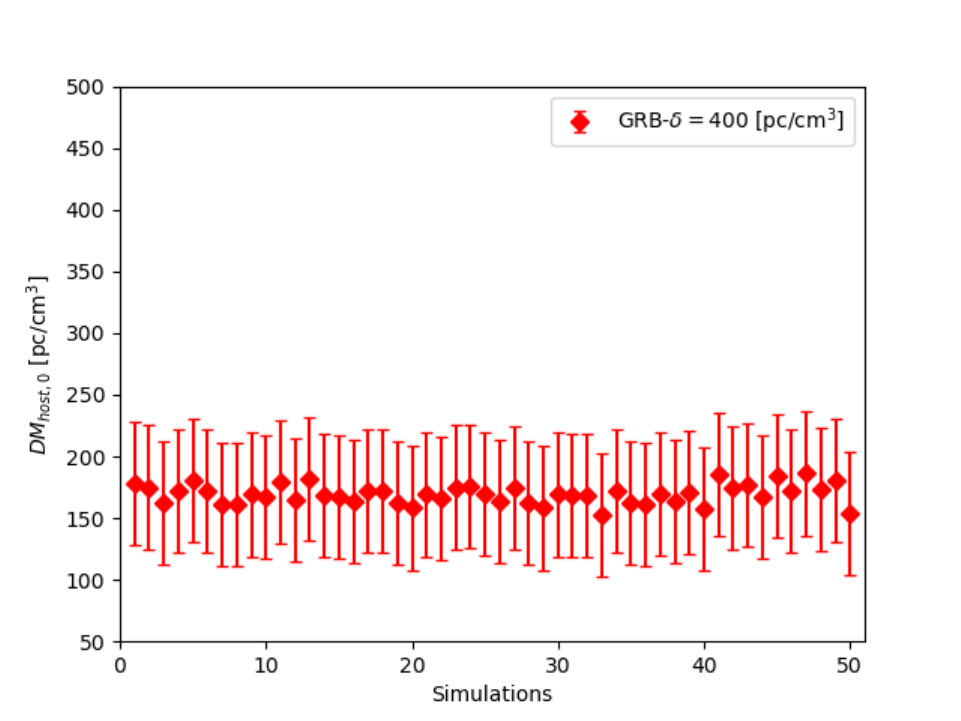}} 
\resizebox{200pt}{130pt}{\includegraphics{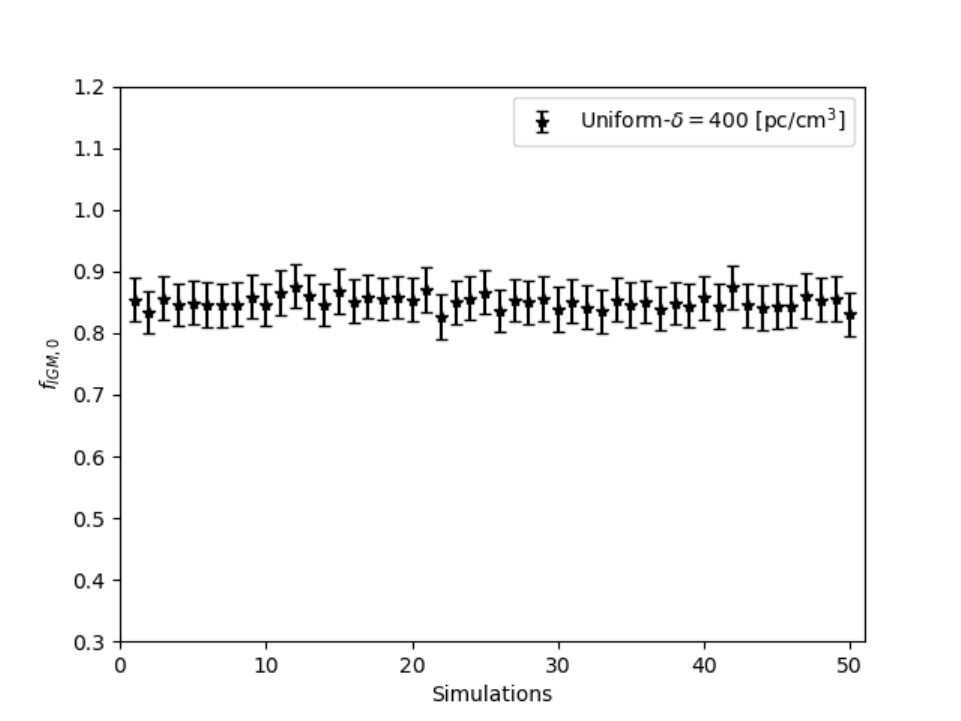}}
\hspace{1mm}\resizebox{200pt}{130pt}{\includegraphics{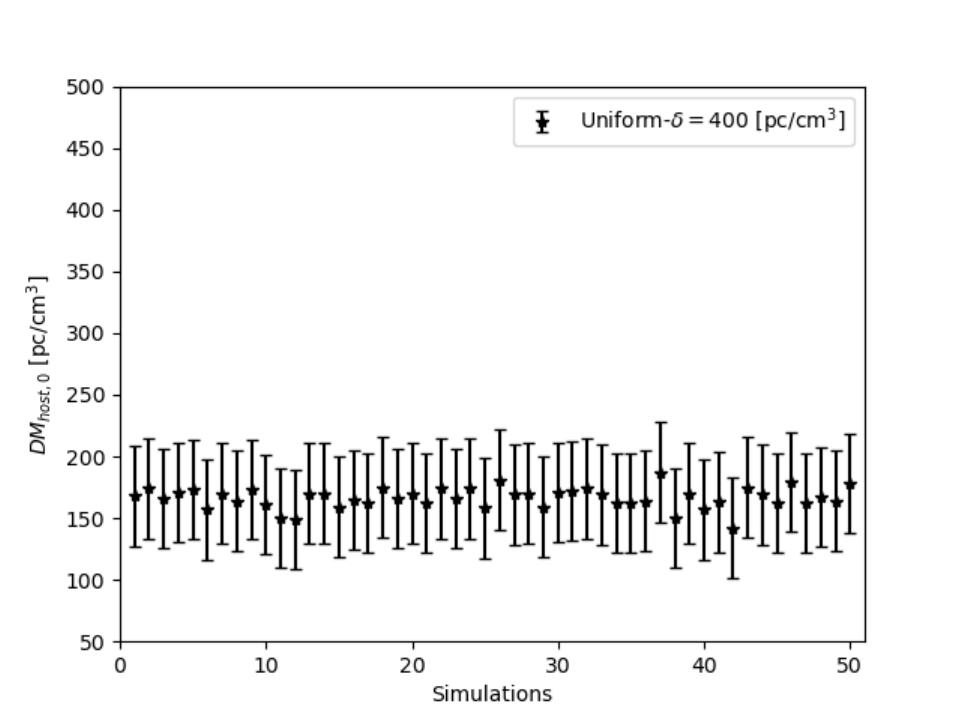}}
\resizebox{200pt}{130pt}{\includegraphics{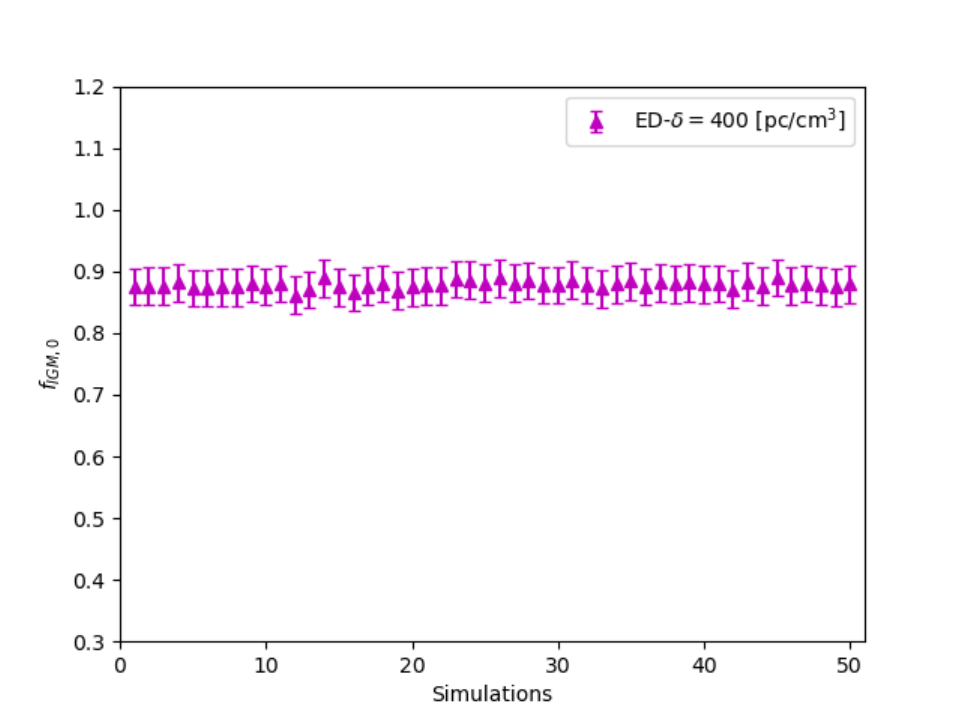}}
\hspace{1mm}\resizebox{200pt}{130pt}{\includegraphics{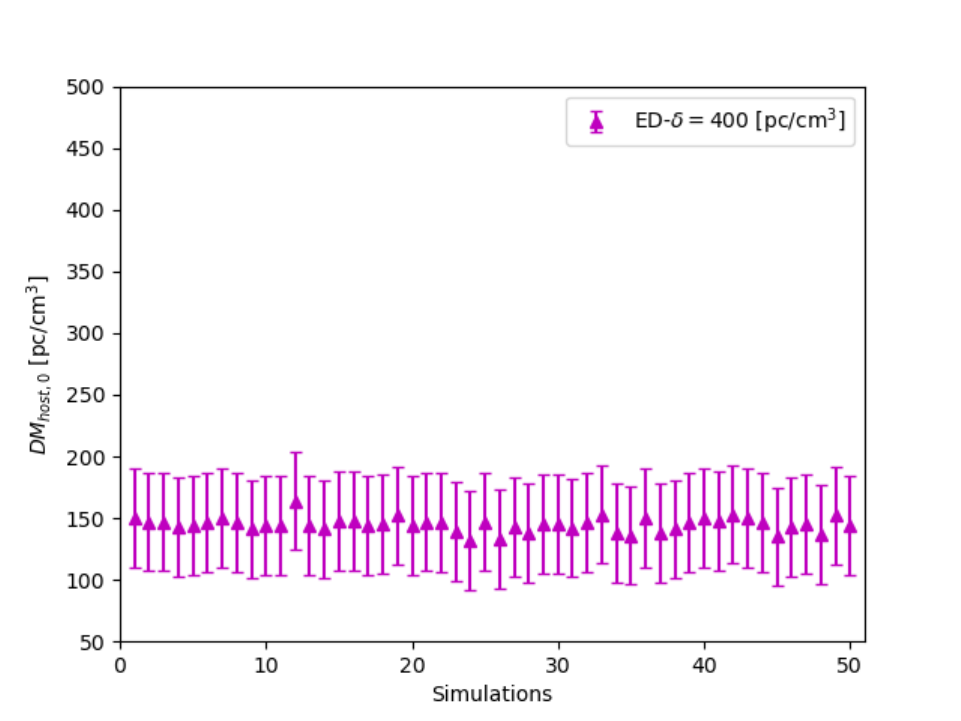}}\\
\end{center}
\caption{The same as in the previous figure, considering $N = 500$ and $\delta = 400$ pc/cm$^{3}$.}  
\label{lowexp1}
\end{figure*}

For completeness, we present below the best-fit results of our simulations for two cases: $N = 15$ and $\delta = 0$ pc/cm$^{3}$ (Fig. \ref{lowexp}) and $N = 500$ and $\delta = 400$ pc/cm$^{3}$ (Fig. \ref{lowexp1}).

\end{document}